\pdfoutput=1
\documentclass[aps,pre]{revtex4}
\usepackage[dvips]{graphics}
\usepackage{graphicx}
\usepackage{amsfonts}
\usepackage{amssymb}
\usepackage{amsmath}
\usepackage{subfigure}
\usepackage{color}

\usepackage{soul}

\begin{document}
\title{Scattering of atomic dark-bright solitons from narrow impurities}

\author{A. \'Alvarez$^{1}$, J. Cuevas$^{2}$, F.R. Romero$^{1}$, C. Hamner$^{3}$, J.J. Chang$^{3}$, P. Engels$^{3}$,
P.G. Kevrekidis$^{4}$, and D.J. Frantzeskakis$^{5}$}

\affiliation{
$^{1}$
Grupo de F{\'i}sica No Lineal, Universidad de
Sevilla. \'Area de F\'{\i}sica Te\'orica. \\
Facultad de F\'{\i}sica, Avda. Reina Mercedes, s/n, E-41012
Sevilla, Spain
\\
$^{2}$  Grupo de F{\'i}sica No Lineal, Universidad de
Sevilla. Departamento de F{\'i}sica Aplicada I. \\ Escuela
Polit{\'{e}}cnica Superior, C/ Virgen de {\'{A}}frica, 7, E-41011
Sevilla, Spain
\\
$^{3}$ Washington State University, Department of Physics and
Astronomy, Pullman, Washington 99164, USA
\\
$^{4}$Department of Mathematics and Statistics, University of
Massachusetts, Amherst MA 01003-4515, USA
\\
$^{5}$ Department of Physics, University of Athens, Panepistimiopolis,
Zografos, Athens 157 84, Greece
}

\begin{abstract}

In this work, we study the collision of an atomic dark-bright soliton, in a two-component Bose-Einstein condensate,
with
%In the present work, we consider a two component Bose-Einstein
%condensate where single dark-bright solitons can appear. We numerically
%investigate the dynamics of these localized structures when they collide with
a Gaussian barrier or well. First, we present the results of
%We start by motivating these investigations through
an experiment, illustrating
the classical particle phenomenology (transmission or reflection)
in the case of an equal barrier or well
in both components. Then, motivated by the experimental observations,
we perform systematic simulations 
%Our theoretical/numerical investigations extend
%past this simplest case, 
considering not only the
case of equal heights, but also
%more importantly
the considerably more complex
setting, where the potential affects only one of the two components.
We systematically classify the ensuing cases within a two-parameter
diagram of barrier amplitudes in the two components, and provide intuitive
explanations for the resulting observations, as well as of their
variations as the size of the barrier changes.
%one affecting the dark field and the other one to the bright field.
%The impurities are modeled through Gaussian functions...
\end{abstract}

%\date{\today}
\maketitle

\section{Introduction}

Atomic Bose-Einstein condensates (BECs) \cite{book2a,book2} provide
an ideal platform for the study of
nonlinear phenomena at the mesoscopic scale (see, e.g., the reviews \cite{emergent,revnonlin,rab,djf}). In
this context, of particular interest are multi-component BECs, which --
in their simplest version -- are
composed of two different hyperfine atomic levels of the same
atom (e.g., $^{87}$Rb \cite{Myatt97,Hall98}).
In such systems, a rich variety of structures can be observed, which
can {\it not} arise in
single-component BECs. Regarding macroscopic nonlinear excitations of
multi-component BECs, a distinctive feature of interest
involves the potential formation of dark-bright (DB) solitons.
This type of vector
soliton consists of
a dark soliton in one component coupled to a bright soliton in the second
component. These
solitons in repulsive BECs are usually referred to as
``symbiotic'' solitons, with this characterization
stemming from the fact that the bright component cannot be supported in a stand-alone fashion (it is only supported as such
in attractive BECs \cite{rab}), and is only sustained because
of the presence of its dark-counterpart, which
acts as an external trapping potential.

%Dark-bright (DB)
Dark-bright solitons have been studied extensively in different settings
in a large number of theoretical works (see, e.g.,
Refs.~\cite{buschanglin,DDB,kanna,rajendran,val,berloff,VB,Alvarez,vaspra,vasnjp}),
while they have also been observed in experiments, both in two-component 
%rubidium 
$^{87}$Rb BECs
\cite{sengdb,peterprl,peter1,peter2,peterpra,peter3} and in nonlinear optics \cite{seg1,seg2,seg3}.
In the BEC context, these experimental studies have chiefly involved the dynamics of a
single DB
%solitary wave
soliton in a trap~\cite{sengdb,peter1}, the generation of
%but also the potential of creating
multiple DB solitons in a counterflow
experiment~\cite{peterprl}, the study of
%and manifesting
their interactions~\cite{peter2}, as well as the creation of SU(2)-rotated DB solitons,
%rotation thereof
in the form of beating dark-dark solitons
%also observable dark-dark solitary waves~
\cite{peterpra,peter3}.

On the other hand, the fundamental problem of the interaction of solitons with localized impurities
has been considered both in nonlinear wave theory \cite{RMP} and solid state physics
\cite{kos1}. 
The interaction of either bright or dark solitons with $\delta$-like
impurities has been investigated in the framework of the nonlinear Schr\"{o}dinger (NLS) equation (see, e.g.,
Refs.~\cite{kos2,cao,holmes,holmer,vvk1}). Relevant studies in the physics of atomic BECs have also appeared some time ago
(see, e.g., Refs.~\cite{gt,nn,greg}), but also more recently, both in the setting of potential wells~\cite{pantofl,pantofl1},
and in that of barriers~\cite{gardiner,martin}; see also the very recent 
work of~\cite{molmer}. 
In this context, localized impurities can be created as
focused far-detuned laser beams and have already been used in experiments with dark solitons
\cite{engels,hulet}; we also note very recent experiments with matter-wave bright solitons of $^7$Li \cite{randy} and
$^{85}$Rb \cite{dic} atoms and localized barriers.
However, such soliton-defect interactions are far less well explored in the case of the multi-component setting
(see, e.g., Ref.~\cite{vaspra} where the statics of DB solitons was studied in the presence of $\delta$-like
impurities). It is the aim of this work to address this problem and study, in particular, the scattering of
atomic DB solitons at narrow impurities.

%However, another dimension which has recently gained considerable
%momentum involves the interaction of solitary waves with defects.
%In the context of BECs, this has been extensively studied
%both in the setting of wells~\cite{pantofl,pantofl1},
%as well as in that of barriers~\cite{gardiner,martin}, in
%part also motivated by recent mathematical~\cite{holmer} advances
%and experimental implementations~\cite{randy} thereof.
%However, such soliton-defect interactions are far less
%well explored in the case of multi-component and especially
%so dark-bright solitary waves and it is that phenomenology
%that we aim at addressing herein.

Our presentation will be structured as follows. In section II, we
present the relevant prototypical model setup in the form of two
coupled Gross-Pitaevskii (GP) equations describing the dynamics of a binary
BEC with repulsive interactions; DB solitons for this model are presented as well.
%defocusing nonlinear Schr{\"o}dinger (NLS) equations.
%Then,
We also present results of a prototypical experiment
where the scattering of DB solitons at a barrier -- which is present (and equal) in both atomic components --
is studied; this experiment provides the motivation for a more systematic theoretical study which is
presented in the next section.
%this experimental motivation involving the case of dark-bright solitons incident
%on a barrier which is present (and equal) in both atomic components.
In particular, in sections III, we numerically explore the dynamics of single DB solitons
in a harmonic trap. The simpler scenario that is studied 
%of
refers to the case where
%concerns
the impurity is equal in the two components -- as in the case of the experiment;
%which, similarly to the experiment, is equal in the two components.
%There,
we find that a particle-based phenomenology is sufficient to
capture the main characteristics here. On the other hand, we identify
a far more significant wealth of possibilities in the setting
where the barrier (or well) is applicable only in one of the two
components. We
%provide
present a systematic study within the plane of
the amplitudes of the two-components, providing intuitive explanations
(on the basis of effective potentials), wherever possible, for the
observed phenomenology.
%with one or two impurities affecting the dark and
%bright fields.
Finally, in section IV, we summarize our findings and
present our conclusions, as well as a number of directions for
potential future studies.

\section{Model and experimental motivation}

\subsection{Gross-Pitaevskii equations and dark-bright solitons}

We consider a two-component
%elongated (along the $x$-direction) repulsive
BEC composed of two different hyperfine states of the same alkali isotope.
If this binary condensate is confined
%of
in a highly anisotropic trap
%(i.e., if the
(with longitudinal and transverse trapping frequencies
%are such that
$\omega_x \ll \omega_{\perp}$), then the mean-field dynamics of the BEC
can be described by the following system of two coupled
%Gross-Pitaevskii (GP)
GP equations
%of the form
\cite{book2a,book2}:
\begin{eqnarray}
i\hbar \partial_t \psi_j =
%\nonumber \\
\left( -\frac{\hbar^2}{2m} \partial_{x}^2 \psi_j +V_j(x) -\mu_j + \sum_{k=1}^2 g_{jk} |\psi_k|^2\right) \psi_j,
\label{model}
\end{eqnarray}
where $\psi_j(x,t)$ ($j=1,2$)
%denote the mean-field
are the macroscopic wave functions of the two components normalized to
the numbers of atoms $N_j = \int_{-\infty}^{+\infty} |\psi_j|^2 dx$, $m$ is the atomic mass,
%and
$\mu_j$ are the chemical potentials,
%furthermore,
$g_{jk}=2\hbar\omega_{\perp} a_{jk}$ are the effective 1D coupling
constants ($a_{jk}$ are
%denote
the
%three
$s$-wave scattering lengths),
%note that $a_{12}=a_{21}$) which account
%for collisions between atoms belonging to the same ($a_{jj}$) or different ($a_{jk}, j \ne k$) species,
while $V_j(x)$ denote the external trapping potentials for each species. In our considerations
below, we will assume that the component $1$ ($2$) supports a dark (bright) soliton; additionally, we
will assume that both components are confined by the usual harmonic trap, namely $V_H(x)=(1/2)m\omega_x^2 x^2$,
while -- for each component -- an additional localized ``impurity'' potential, which
can be generated by off-resonant Gaussian laser beams, is also present. Thus, the external potentials
$V_j(x)$ for each of the two components
%can be
are described as:
\begin{equation}
V_1(x)=V_H+E_d\exp\left(-\frac{2x^2}{\epsilon_d^2}\right),
\quad
V_2(x)=V_H+E_b\exp\left(-\frac{2x^2}{\epsilon_b^2}\right),
%\quad
%V_0(x)=\frac{1}{2}m \omega_x^2 x^2
\end{equation}
where the parameters $E_d$, $E_b$ and $\epsilon_d$, $\epsilon_b$ set, respectively, the amplitudes and widths of the %barriers
impurities in each component. Notice that for a blue- or red-detuned laser beam, the impurity potentials can either {\it repel} ($E_{d,b}>0$) or {\it attract} ($E_{d,b}<0$) the atoms of the respective component of the condensate.

%\jcm{\bf JCM: Some of the relations below should be removed. We must take into account that the energies in the simulations are given in dimensional units (Hz), so any of you (probably Dimitri) should pay a lot of attention to the normalizations, scales, \ldots}.

We now cast Eqs.~(\ref{model}) into a dimensionless form as follows: measuring the densities $|\psi_j|^2$, length, time and energy in units of $2a_{11}$, $a_{\perp} = \sqrt{\hbar/\omega_{\perp}}$, $\omega_{\perp}^{-1}$ and $\hbar\omega_{\perp}$, respectively, Eqs.~(\ref{model}) are reduced to the
%following
form:
\begin{eqnarray}
i \partial_t u  &=& -\frac{1}{2} \partial_{x}^2 u  + V_1(x)u
%\nonumber \\
+(|u|^2 + \tilde{g}_{12}|v|^2 -\mu_1) u,
\label{deq1}
\\
i\partial_t v  &=& -\frac{1}{2} \partial_{x}^2 v +V_2(x) v
%\nonumber \\
+ (\tilde{g}_{12}|u|^2 + \tilde{g}_{22} |v|^2 - \mu_2) v.
\label{deq2}
\end{eqnarray}
In Eqs.~(\ref{deq1})-(\ref{deq2}), the wave functions $u$ and $v$ correspond to $\psi_1$ and $\psi_2$
respectively, the normalized nonlinearity coefficients are given by $\tilde{g}_{j2} = g_{j2}/g_{11}$,
%($j=1,2$),
while the normalized harmonic trap potential (incorporated in $V_1(x)$ and $V_2(x)$ as discussed above) is now
given by $V_H(x)=(1/2) \Omega^2 x^2$, where $\Omega = \omega_x/\omega_{\perp}$.

Notice that in the GP Eqs.~(\ref{deq1})-(\ref{deq2}) the number of atoms $N_{D,B}$ of each component
is conserved; in fact, $N_{D,B}$ are given by $N_{D,B} = (a_{\perp}/2 a_{11}) \tilde{N}_{D,B}$, where
%
%\begin{equation}
%    \tilde{N}_D=\int_{-\infty}^{\infty}|u(x)|^2\,\mathrm{d}x ,
%    \qquad
%    \tilde{N}_B= \int_{-\infty}^{\infty}|v(x)|^2\,\mathrm{d}x,
%\end{equation}
%
$\tilde{N}_D=\int_{-\infty}^{\infty}|u(x)|^2\,\mathrm{d}x$ and
$\tilde{N}_B= \int_{-\infty}^{\infty}|v(x)|^2\,\mathrm{d}x$
are the respective integrals of motion of the normalized GP Eqs.~(\ref{deq1})-(\ref{deq2}).
%Additionally, the total energy of the system (Hamiltonian) $E$ is conserved; this quantity is given by
%$E= \hbar \omega_{\perp} \tilde{E}$, where
%
%\begin{eqnarray}
%\tilde{E} &=& \frac{1}{2}\int_{-\infty}^{+\infty} \mathcal{E} dx, \nonumber \\
%\mathcal{E} &=& |\partial_{x} u|^2+|\partial_{x} v|^2 + |u|^4+\tilde{g}_{22}|v|^4
%+ 2 \tilde{g}_{12}|u|^2 |v|^2 + 2 [V_1(x)-\mu_1]|u|^2 + 2 [V_2(x)-\mu_2]|v|^2,
%\label{energy}
%\end{eqnarray}
%
%is the respective Hamiltonian of the normalized GP Eqs.~(\ref{deq1})-(\ref{deq2}).

As mentioned above, in the physically relevant setting of $^{87}$Rb, the scattering lengths characterizing
the intra- and inter-component atomic collisions are almost equal; thus, to a first approximation,
one may assume that $\tilde{g}_{12} = \tilde{g}_{22} \approx 1$, which means that the system of
Eqs.~(\ref{deq1})-(\ref{deq2}) is of the Manakov type \cite{Manakov}; in this case, the system is integrable
in the absence of the external potentials $V_{1,2}(x)$ and admits exact analytical dark-bright  soliton solutions.
Particularly, considering the boundary conditions
$|u|^2\rightarrow \mu_1$ and $|v|^2\rightarrow 0$ as $|x| \rightarrow\infty$, Eqs.~(\ref{deq1})-(\ref{deq2})
possess an exact analytical one-DB-soliton solution of the following form (see, e.g., Ref.~\cite{buschanglin}):
\begin{eqnarray}
u_{\rm DB}(x,t) &=& \sqrt{\mu_1} \{\cos\phi {\rm tanh} \xi+i \sin\phi \},
\label{db_u_d}
\\
v_{\rm DB}(x,t) &=& \eta {\rm sech}\xi \exp[ikx+i\theta(t)],
\label{db_u_b}
\end{eqnarray}
where $\xi=D(x-x_0(t))$, $\phi$ is the dark soliton's phase angle, $\cos\phi$ and $\eta$ represent the
amplitudes of the dark and bright solitons, and $D$ and $x_0(t)$ are associated with the inverse width
and the center position of the DB soliton. Furthermore, $k=D \rm tan\phi={\rm const}$ and $\theta(t)$
are the wavenumber and phase of the bright soliton, respectively. The above parameters of the DB-soliton
are connected through the
%following
equations: $D^2 = \mu_1 \cos^2\phi-\eta^2$, $\dot{x}_0 = D {\rm tan}\phi$, and
$\theta(t) = (1/2)(D^2-k^2 + \mu_2-\mu_1)t$,
%
%\begin{eqnarray}
%D^2 &=& \mu_1 \cos^2\phi-\eta^2,
%\label{D}
%\\
%\dot{x_0} &=& D {\rm tan}\phi,
%\label{d_x_0}
%\\
%\theta(t) &=& \frac{1}{2}(D^2-k^2 + \mu_2-\mu_1)t,
%\label{theta}
%\end{eqnarray}
%
%with $\dot{x}_0$ denoting
%
where $\dot{x}_0$ is the DB soliton velocity.

%Notice that the amplitude $\eta$ of the bright soliton, the chemical potential $\mu$ of the dark soliton, as well as the (inverse) width parameter $D$ of the DB soliton are connected to the number of atoms $N_{\rm B}$ of the bright soliton by means of the following equation:
%
%\begin{eqnarray}
%\tilde{N}_B =\frac{2 \eta^2}{D}.
%\label{nb}
%\end{eqnarray}
%

%\jcm{\bf JCM: maybe the paragraph about should be removed, as we don't make use of the analytical approximations in the paper}.

%\jcm{\bf JCM: the paragraph below should be merged with the experimental section, as there are some redundant information. Maybe the experimental section could be before this one}.

\subsection{Experimental results}
%Motivation}
\label{sec:experiments}

Having introduced our setup, we now proceed by presenting results of an experiment
dealing with scattering of atomic DB solitons at barriers. In fact, the results that will be
presented below,
%In order to
motivate the more detailed theoretical investigation of this paper, but also
%to
illustrate the experimental tractability of this direction, and
%also to
showcase
prototypical results along this vein.

\begin{figure}
\includegraphics[width=0.9\textwidth]{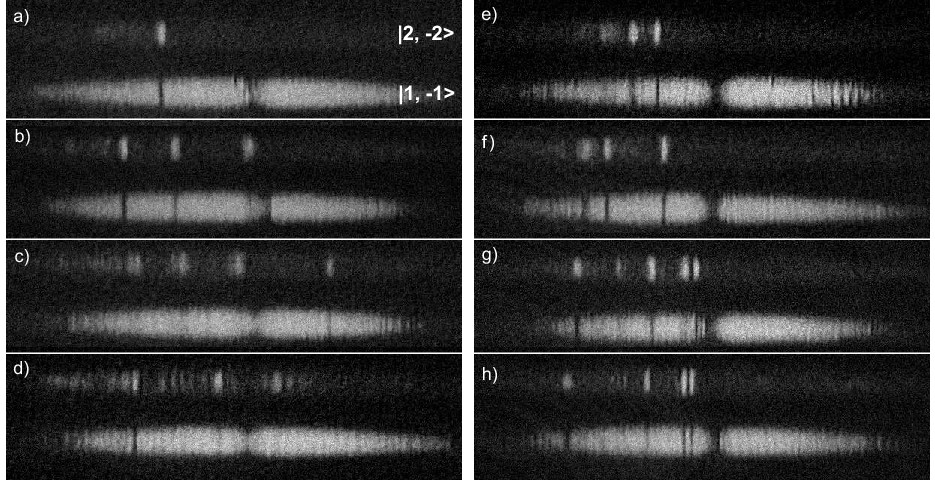}
%\textwidth
% \leavevmode \epsfxsize=3.5in
%  \epsffile{experimentfig.jpg}
\caption{\label{experiment}
(a-d) Time evolution of the soliton-barrier interaction. The peak
barrier potential is $0.56\cdot\mu_{BEC}$. The in-trap evolution
times after generation of the solitons are a) 250~ms, b) 500~ms, c)
750~ms, d) 1000~ms. (e-h) Similar to (a-d) but for a peak barrier
potential of $1.11\cdot\mu_{BEC}$ and evolution times e) 300~ms, f)
600~ms, g) 700~ms, h) 1000~ms. In all cases the chemical potential
of the BEC, $\mu_{BEC}$, is approximately 36~nK (about 750~Hz).}
\end{figure}

Our experimental results are summarized in Fig.~\ref{experiment}.
There, it is shown that,
%presents experimental results of dark-bright solitons that,
depending on the barrier height, DB solitons are either reflected
by or transmitted through a repulsive barrier. The experiment is
conducted with a BEC of $4.5\cdot10^5$ atoms of $^{87}$Rb confined
in an optical dipole trap with trapping frequencies
$\{\omega_{\rm axial}, \omega_{\rm vertical}, \omega_{\rm horizontal}\}=2\pi
%\cdot
\times \{1.4, 120, 174\}$~Hz. The solitons
are generated by transferring a small fraction of the atoms from the
initial $|F,m_{F}\rangle$ = $|1,-1\rangle$ hyperfine state to the
$|2,-2\rangle$ state and exploiting a counter-flow induced
modulational instability \cite{peterprl} generated by a magnetic
gradient along the axial direction. The number of solitons, as well
as their initial positions, can be controlled by adjusting
experimental parameters such as the number of atoms transferred into
the second state and the strength and duration of the magnetic
gradient used to induce the counter-flow. For the data presented in
Fig.~\ref{experiment}, the solitons are generated in the left part
of the BEC, the magnetic gradient is subsequently turned off, and
the solitons start moving towards the trap center. The oscillations
of individual solitons in a trap have been investigated in detail in
\cite{peter1}. For the present data, we additionally ramp on
a repulsive barrier at the center of the trap. The barrier is
generated from a 660~nm laser beam with a narrow waist of
approximately $18$~$\mu$m in the direction of the BEC axis and has an
aspect ratio of 4. For imaging, the two components of the BEC are
vertically separated during a short free expansion time of 7~ms for
the upper cloud and 8~ms for the lower cloud
\cite{peterprl,peter1}. For barrier depths larger than the
chemical potential [cf. Fig.~\ref{experiment}(e-h)] we observe
confinement of the dark-bright solitons to the left half of the BEC.
This is consistent with having two isolated BECs. For a barrier
depth of approximately half the chemical potential
[cf. Fig.~\ref{experiment}(a-d)] we observe solitons penetrating through the
barrier; see, e.g., especially the panel (c) in this setting.
The dynamics observed here is a subset of the rich
behavior expected for soliton-barrier interactions. These dynamics
can be extended to more exotic regimes, e.g., by the addition of a
species selective barrier. The latter will be examined in more
detail in our theoretical investigation below.

\section{Numerical investigations}

%\subsection{Collision simulations of dark-bright solitons when the impurities are identical }\label{sec:identical}

%In the next two sections
%Starting from this section, we will present a survey of the collision
%simulations of
%dark-bright
%DB solitons with impurities, taking $E_b$ and $E_d$ as varying parameters and a fixed value of the barrier
%width $\epsilon$.

%%%%%%%%%%%%%%%%

In our numerical simulations below, we will assume that the two-component BEC under consideration consists
of two different hyperfine states of $^{87}$Rb, namely
%such as
the states $|1,-1\rangle$ and $|2,-2\rangle$ used in the experiment presented in the previous section
(see also Refs.~\cite{sengdb,peterprl,peter1,peter2,peterpra,peter3}). In this case, the scattering
lengths take the values $a_{11}=100.4a_0$, $a_{12}=98.98a_0$ and $a_{22}=98.98a_0$ (where $a_0$ is the Bohr radius);
accordingly, the normalized nonlinearity coefficients in Eqs.~(\ref{deq1})-(\ref{deq2}) take equal values:
$\tilde{g}_{12} = \tilde{g}_{22} \approx 0.986$. Furthermore, we will assume that the trap frequencies are
$\omega_\perp=2\pi\times 116$~Hz and $\omega_z=2\pi\times 1.3$~Hz, i.e., $\Omega \approx 0.0112$, and the
numbers of the atoms in each component are $N_D=70,000$ and $N_B=1,000$ resulting in a chemical potential of approx. 305 Hz for the total atom number. These 
values are similar to the
respective ones used in experiments \cite{sengdb,peterprl,peter1,peter2,peterpra,peter3}.

Concerning the parameters of the Gaussian impurity potential, the values of $E_{b,d}$
are taken in the interval $[-200,200]$, and we fix the value $\epsilon_d=\epsilon_b=\epsilon=3\ \mu$m. Notice that we focus here on relatively narrow
impurities, as for those we have 
explored the steady state problem~\cite{vaspra} and, as will be seen
below, they already present a rich phenomenology. An examination of the
effect of the width of the impurity will be deferred to a future
study.
Our principal aim in what follows is to study the scattering of
%dark-bright
solitons at the impurity potential. To do so, we displace the solitons from the trap center, using the initial position
value $x_0=-40\ \mu$m, which is sufficiently far from $x=0$, so that the solitons do not overlap
with the impurity. We then ``release'' the solitons and observe
their subsequent interaction with the Gaussian barrier, measuring
the fraction and observing also the shape of the atoms that are
transmitted, reflected, and trapped at the potential.
We will
%assume that $E_b$ and $E_d$ are varying parameters as mentioned above and
study, at first, the case $E_b=E_d$, as per our experimental results
(cf. 
%sec.~\ref{sec:identical}), 
subsection A below) and then the case where the 
impurity acts only on one component,
i.e., either $E_d=0,~E_b\ne0$ or $E_d\ne 0,~E_b=0$ (cf. 
%sec.~\ref{sec:no identical}).
subsection B below).

\begin{figure}[h]
\begin{center}
\includegraphics[width=0.6\textwidth]{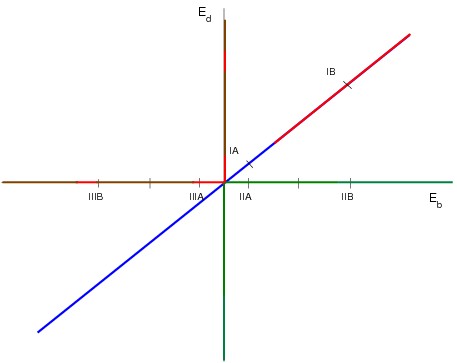}
\end{center}
\caption{(Color online) A state diagram in the parameter space ($E_b,E_d$). The
different regimes that appear after the collisions are illustrated by
colors: (Blue) Transmission regime. (Red) Reflection regime. (Green)
Transmission-reflection regime. (Brown)
Trapping-transmission-reflection regime. $A$ and $B$ correspond,
respectively, to a small and large value of the parameters in each
region: I ($E_d=E_b=E$), II ($E_d=0, E_b>0$) and III ($E_d=0,
E_b<0$).} \label{Fig1}
\end{figure}

At this point, it is relevant to present
%We start by presenting %Fig.~\ref{Fig1} shows
a sketch of a state diagram in the parameter
space $(E_b,E_d)$, as depicted in Fig.~\ref{Fig1}. The different regimes that appear after the
collisions are illustrated by colors and they will be discussed
below. The capital letters $A$ and $B$ correspond, respectively, to
a small and large value of the parameters in each region;
the cases I ($E_d=E_b=E$), II ($E_d=0,~E_b>0$) and III ($E_d=0,~E_b<0$) correspond
to the principal cases that we will examine in what follows. For
each $A$ and $B$ we will illustrate the contour plots of the
densities of both components.
 Furthermore, in the case where $E_d=E_b=E$ where as we will see
below the ``particle-like'' picture is most relevant, we will also
display an effective potential energy landscape encountered by the 
DB solitons. This amounts to computing the turning point, say $x_1$, of each of
our initializations of the DB soliton at position $x_0$ 
(whose potential energy $V_{H}(x_0)$ in a harmonic
oscillator trap we can evaluate). Then, $(x_1,V_{H}(x_0))$ is identified
as a point in the effective potential energy surface.
%that the solitary waves encounter.

%{\bf Jesus: we should write sth. here about how this is practically
%computed.}

\subsection{Scattering of DB solitons from identical impurities}
%Collision simulations of dark-bright solitons when the impurities are identical }
\label{sec:identical}

We start with the case where both impurities are
identical, i.e.,
%that is
$E_d=E_b=E$, either repulsive ($E>0$) or attractive ($E<0$).

In the repulsive case of $E_d=E_b=E>0$, the simulations reveal
%markedly demonstrate
the existence of two different regimes.
For sufficiently small values of the repulsive barrier $E$, the
%dark-bright
DB solitons are transmitted (transmission regime, blue
color in region I of Fig.~\ref{Fig1}).
%When the potential energy of the DB-soliton particle
%is larger than the height of the effective barrier encountered
%by it (measured as the energy
%of the DB soliton at the center with minus that without the
%barrier),
For values of $E$ bigger than
a critical value, i.e., $E\gtrsim20$~Hz
%\jcm{
(for solitons launched from $x_0=40\ \mu$m)
%},
the DB solitons are reflected
(reflection regime, red color in the same region I of Fig.~\ref{Fig1}).
Thus, in this case, the solitons behave as classical particles: if they have potential
energy (recall that the solitons start with zero kinetic energy) larger than the height of the barrier
then they are transmitted through it, while they are reflected in the opposite case.

\begin{figure}[h]
\begin{center}
\includegraphics[width=0.6\textwidth]{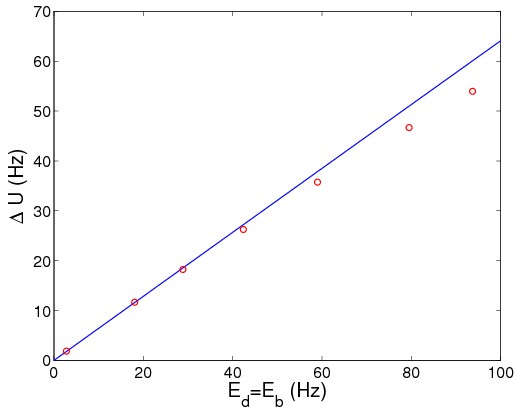}
\end{center}
\caption{(Color online) A comparison of the effective potential energy
barrier $\Delta U$ encountered by the DB soliton (solid line)
with the
%with the
critical energy of the soliton
%solitary wave
that separates the transmission from the reflection regime, for different values
of $E_d=E_b=E$ (red circles). The very good agreement (especially
%so
for smaller values $E$) lends support to the consideration of this case
as an example of a classical (solitonic) particle scattering from a barrier.}
\label{fig_hdb1}
\end{figure}

This particle-like behavior can be quantitatively described as follows. First, we consider
%\jcm{T
the maximum of the effective barrier height that the DB-soliton particle encounters at the defect,
which can be defined as:
\begin{equation}
    \Delta U=U_\mathrm{def}-U_0,
\end{equation}
with $U_\mathrm{def}$ ($U_0$) being the normalized potential energy of
the DB soliton at the
%bottom of the
trap center with(out) the defect:
\begin{equation}
    U_\mathrm{def}=\frac{\int_{-\infty}^\infty\,\mathrm{d}x\,V_1(x)|u'(x)|^2 dx}{\int_{-\infty}^\infty\,\mathrm{d}x\,|u'(x)|^2}+
    \frac{\int_{-\infty}^\infty\,\mathrm{d}x\,V_2(x)|v'(x)|^2 dx}{\int_{-\infty}^\infty\,\mathrm{d}x\,|v'(x)|^2},
\end{equation}
\begin{equation}
    U_0=\frac{\int_{-\infty}^\infty\,\mathrm{d}x\,V_0(x)|u_0(x)|^2 dx}{\int_{-\infty}^\infty\,\mathrm{d}x\,|u_0(x)|^2}+
    \frac{\int_{-\infty}^\infty\,\mathrm{d}x\,V_0(x)|v_0(x)|^2 dx}{\int_{-\infty}^\infty\,\mathrm{d}x\,|v_0(x)|^2}.
\end{equation}
%
%and with
In the above expressions, $\{u'(x),v'(x)\}$ and $\{u_0(x),v_0(x)\}$
%being
denote, respectively, the DB solitons at the trap center with and without the defect,
as found numerically
%computed
by means of a fixed point algorithm, using as an initial guess Eqs.~(\ref{db_u_d})-(\ref{db_u_b}).
%at the %bottom of the trap center with(out) the defect.
Notice that the solutions with the prime are found by keeping fixed the chemical potentials $\mu_{1,2}$ of each component to those of the solution without defect.
%
%This quantity is equivalent to the BEC chemical potential $\mu_\mathrm{BEC}$ defined at Section \ref{sec:experiments} %\textbf{\ldots if I'm not wrong}
%}.

%\jcm{\bf JCM: Dimitri, why we should divide by the norm in the potential energy but not at the total energy?}

%\jcm{
Figure~\ref{fig_hdb1} shows that the critical value separating the reflection and transmission regime
(above and below the solid line in the figure) is very accurately obtained from the comparison of the potential energy of
the DB-soliton particle with $\Delta U$. Note that in the simulations (see red circles in the figure)
we fix the initial soliton location $x_0$ and vary $E_d=E_b$ in order to determine the critical value which separates the
reflection and transmission regimes for this value of $x_0$ or, equivalently, of the potential energy of the soliton,
i.e., $(1/2) \Omega^2 x_0^2$.
%
%This quantity is measured as the effective potential energy of the bright part of the
%solitary wave
%soliton inside the parabolic trap (i.e., as $1/2 \omega_x^2 x_0^2$ for the critical $x_0$ that separates the different regimes).

%This critical value is very accurately obtained
%as shown in Fig.~\ref{fig_hdb1} from the comparison
%of the potential energy of the DB-soliton particle
%and that of the effective barrier that it encounters
%at the defect.

%The latter ($\Delta U$) is measured
%as the energy
%of the DB soliton at the center in the presence of the
%paper minus that in the absence of the
%barrier. The former is measured as the effective potential
%energy of the bright part of the solitary wave inside
%the parabolic trap (i.e. as $1/2 \omega_x^2 x_0^2$ for
%the critical $x_0$ that separates the different regimes).

\begin{figure}[h]
\begin{center}
\begin{tabular}{cc}
\multicolumn{2}{c}{\includegraphics[width=0.7\textwidth]{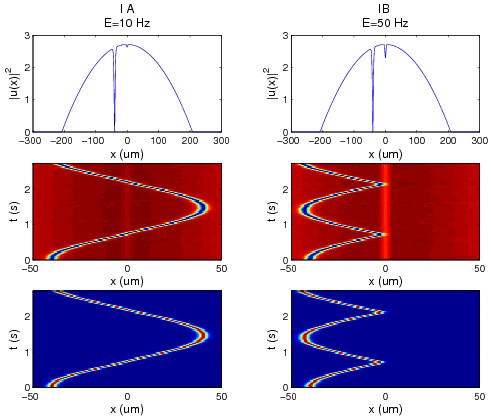}} \\
\includegraphics[width=0.35\textwidth]{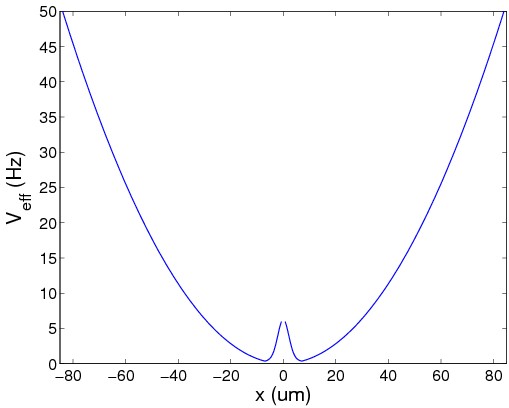} &
\includegraphics[width=0.35\textwidth]{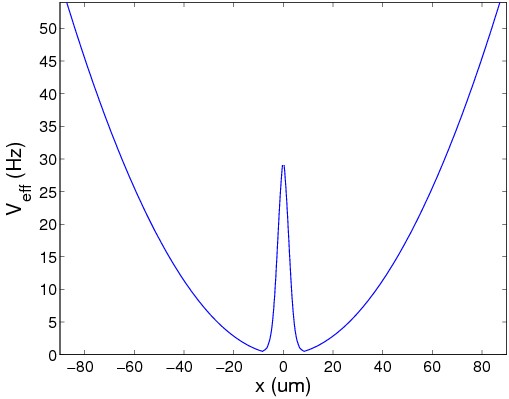}
\end{tabular}
\end{center}
\caption{(Color online) The case of identical repulsive impurities. Top row: initial density
profiles of the dark components for a small and large value of
$E>0$. Middle rows: contour plots of the densities of the dark
components $|u(x)|^2$
%. Bottom row: contour plots 
and of the
densities of the bright components $|v(x)|^2$.
The left panels correspond to a subcritical case, featuring full
(nearly full) transmission, while the right ones are for a supercritical
case with (nearly complete) reflection.
Bottom panels: the effective potential encountered by the solitary wave.
The low height of the barrier in the left column enables transmission,
while its increase in the right panel induces the reflection.
\label{Fig2}}
\end{figure}

The contour plots of the densities of the dark and bright components
corresponding to relatively small and large values of $E>0$ (corresponding to $IA$ and $IB$
%respectively, %of
in Fig.~\ref{Fig1}) are illustrated in Fig.~\ref{Fig2}
(columns $IA$ and $IB$). The top row shows the initial density
profiles of the corresponding dark components. It is important to highlight
here (as it will become also relevant for other cases) that the dark
component sustains an increasing density dip as $E$ increases in
positive values, while it will correspondingly feature a density
bump in the case of increasing negative such values. This is a feature
of the ground state in the presence of the defect, as the latter
repels for $E>0$ and attracts for $E<0$ the atoms in the neighborhood
of $x=0$. It is clear also by plotting the effective potentials
(in the bottom panel of the figure) that the DB faces a weak barrier
in the former case and its potential energy is sufficient to overcome
it. On the other hand, the barrier is considerably higher in the right
panel, inducing the reflection of the solitary wave.

We complete this section by noting that in the case where both impurities are attractive, i.e., $E_d= E_b=E<0$, the
%dark-bright
DB solitons are always transmitted after the collision, for
every value of $E$,
% value,
hence there is only a transmission regime depicted by blue color in Fig.~\ref{Fig1}.

\subsection{Scattering of DB solitons from an impurity in one component}
%{Collision simulations of dark-bright solitons with an impurity in only one component}
\label{sec:no identical}

While the dynamical evolution of the case where the defect acts on
%affects
both components was found to be fairly straightforward, the
case where the barrier is imposed selectively on only one of
the components was found to be considerably more complex.
We considered both the case where $E_d=0$ and $E_b\neq 0$ and that
where $E_d\neq 0$ and $E_b=0$. The results show that the following  two subcases are equivalent:

\vspace{0.5cm}

a) $(E_d=0, E_b>0) \equiv (E_d<0, E_b=0)$.

\vspace{0.5cm}

b) $(E_d=0, E_b<0) \equiv (E_d>0, E_b=0)$.

\vspace{0.5cm}

 The case where $E_d=0$ and $E_b<0$ represents the existence of an attractive
barrier in the bright component and absence of impurity in the dark
component. The above equivalence can be most easily qualitatively
appreciated in that case b), hence we present it for that setting. In
particular,
when an impurity attractively affects the atoms in the bright component,
then it favors the ``collection'' of atoms near the origin.
This, in turn, builds a population of atoms in that neighborhood
which, through the term proportional to $g_{12}$ in the equations
of motion, provides a repulsive barrier for the dark component.
Hence, the existence of an attractive well solely in the
bright component becomes tantamount to having a repulsive barrier
in the dark component. An analogous argument can be used
to showcase that a repulsive barrier in the bright component,
through favoring the absence of atoms in its vicinity, creates
an effective well for the dark component atoms.  The above
feature is directly evident in the diagram of Fig.~\ref{Fig1},
hence we only focus on each of the representatives of the cases
a) and b) above.

\begin{figure}[tp]
\begin{center}
\begin{tabular}{cc}
\includegraphics[width=0.35\textwidth]{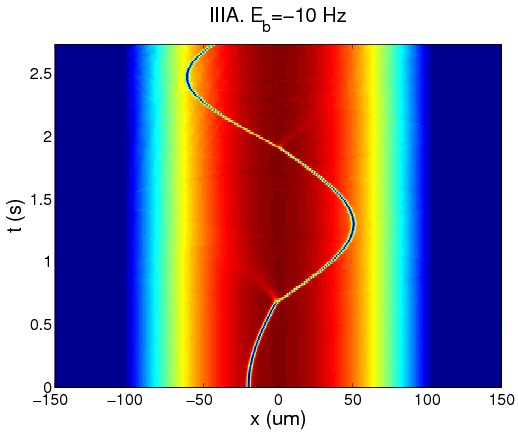} &
\includegraphics[width=0.35\textwidth]{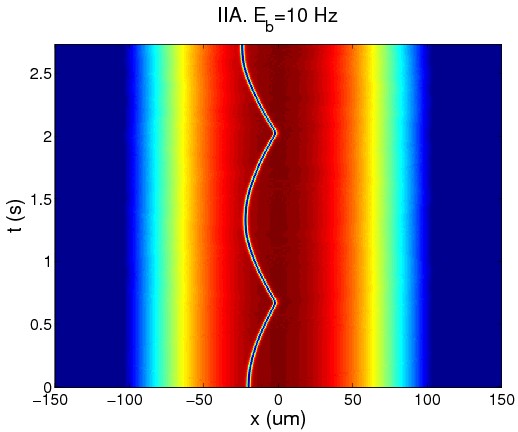} \\
\includegraphics[width=0.35\textwidth]{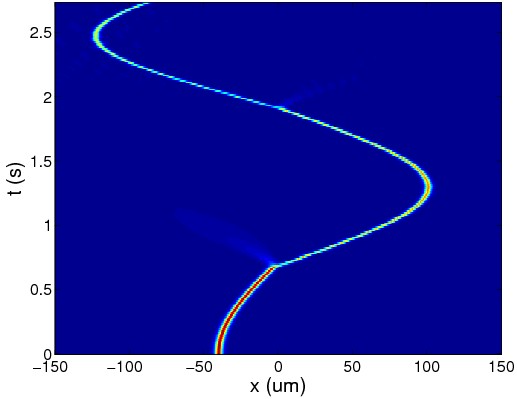} &
\includegraphics[width=0.35\textwidth]{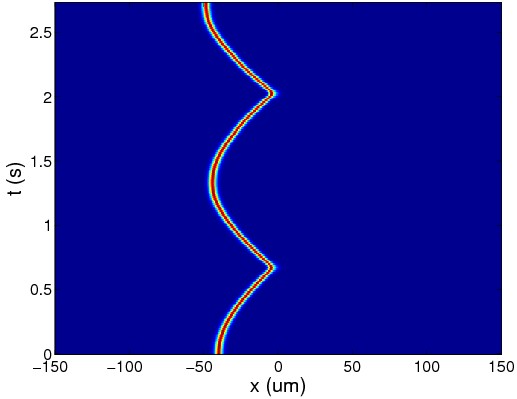}
\end{tabular}
\end{center}
\caption{(Color online) Comparison, for a small value of $|E_b|$, between the case
with repulsive bright impurity and the case with attractive bright
impurity, with $E_d=0$. Column IIA: repulsive bright
impurity with $E_b=10$Hz and $E_d=0$. Column IIIA: attractive bright
impurity with $E_b=-10$Hz and $E_d=0$. 
%Top row: effective potentials.
Top row: the dark components $|u(x,t)|^2$. Bottom row: the
bright components $|v(x,t)|^2$.} \label{Fig3}
\end{figure}

After the collisions, for $E_d=0, E_b>0$,
part of the energy is transmitted
and part of it is reflected. We denote this as a
transmission-reflection
regime (green color in Fig.~\ref{Fig1}). For small values of $E_b$
the
%dark-bright
DB solitons are mainly transmitted, and when $E_b$ is
high enough they are mostly reflected. Equivalent results, when the
well depth $|E_d|$ increases, are obtained for the case where
$E_d<0$ and $E_b=0$, which represents the existence of an attractive
well in the dark component and the absence of impurity in the bright
one. For small $E_b>0$, this dynamics
%as shown in the left
%panel of Fig.~\ref{Fig3}, 
can be understood 
%on the basis of
%effective dynamics. 
as follows. As discussed above, the repulsion of bright
atoms produces an effective attraction of dark atoms, hence
inducing an effective potential well, rather than
what was anticipated as a potential barrier. It should be noted
here that this counter-intuitive effect was quantified in the case of
a $\delta$-function potential in Ref.~\cite{vaspra}. 
%It is illustrated
%qualitatively here, as well, in the top left panel of Fig.~\ref{Fig3},
%where it is evident that $E_b>0$ leads to an effective potential
%well. 
This effect leads to the acceleration of the soliton
%wave
(with a small back-scatter due to the inelasticity of collision
with the defect) visible in the left panels
of the figure. To complete the discussion of Fig.~\ref{Fig3},
let us briefly touch upon the right panels of the figure.
This concerns the case of $E_b<0$ (while $E_d=0$, namely case
b) above). The corresponding situation here, when $E_b$ is
small presents a repulsive effect for the dark atoms and as such
results in an effective barrier.
% mirrored in the top right panel
%of the figure. 
This prediction is also corroborated by the
analytical considerations for the $\delta$-function case
of Ref.~\cite{vaspra}. This, in turn, leads to the reflection
dynamics observed in the 
%middle and bottom 
right panels
of Fig.~\ref{Fig3}.

%\vspace{1cm}

% b) $(E_d=0, E_b<0) \equiv (E_d>0, E_b=0)$.

%\vspace{0.7cm}

\begin{figure}[tp]
\begin{center}
\begin{tabular}{cc}
\includegraphics[width=0.35\textwidth]{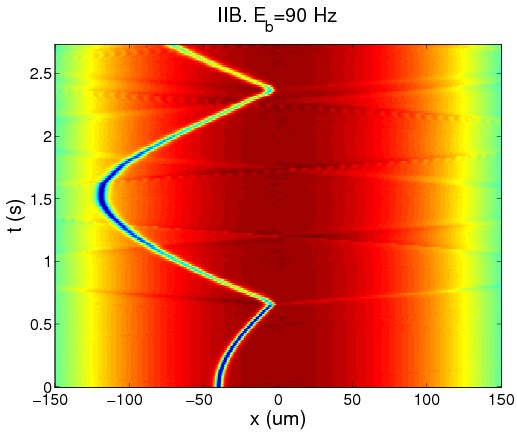} &
\includegraphics[width=0.35\textwidth]{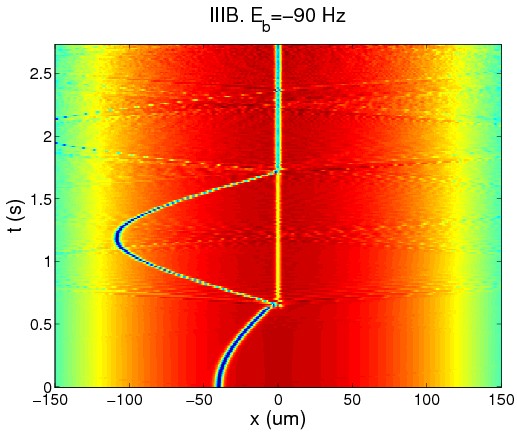} \\
\includegraphics[width=0.35\textwidth]{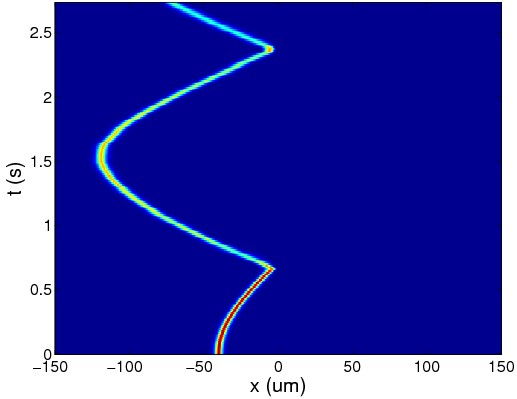} &
\includegraphics[width=0.35\textwidth]{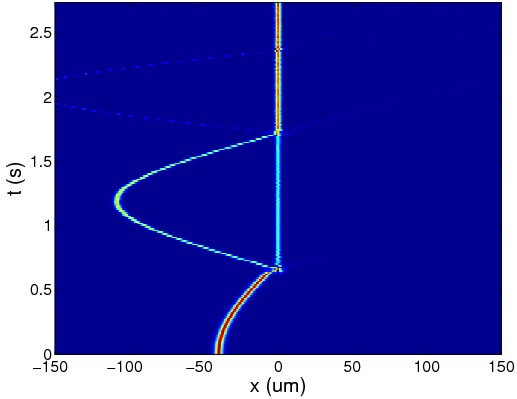}
\end{tabular}
\end{center}
\caption{(Color online) Comparison, for a high value of $|E_b|$, between the case
with repulsive bright impurity and the case with attractive bright
impurity, with $E_d=0$. Column IIB: repulsive bright
impurity with $E_b=90$Hz and $E_d=0$. Column IIIA: attractive bright
impurity with $E_b=-90$Hz and $E_d=0$. 
%Top row: effective potentials.
Top row: the dark components $|u(x,t)|^2$. Bottom row: the
bright components $|v(x,t)|^2$.
%\jcm{
%Notice that the reflection point in the barrier is farther from the defect when% increasing the initial soliton velocity, contrary to the case $E_b=E_d$.
} \label{Fig4}
\end{figure}

We now turn to the case of large barrier strength in Fig.~\ref{Fig4}.
In this setting, there is a fundamental difference in comparison
to the case of weak barrier presented previously. This consists
of the fact that for small $E_b$, the defining characteristic
in the DB-soliton and defect interaction is the nature of the
potential for the dark component (which, as we saw, was
somewhat counter-intuitively the opposite than the one for
the bright component). However, for large $E_b$, the nature
of the potential for the bright component becomes
important and hence in this case, large positive $E_b$
also induces a locally strong repulsive potential for the bright atoms.
On the other hand, large negative $E_b$ creates a large
attractive potential for the bright atoms. However, the latter tends to favor
the trapping of the atoms of the bright component
%atoms
together with those
of the dark component, leading essentially to the formation
of a defect mode, alongside a partial reflection of the
soliton. 
%(in analogy with bottom right panel of Fig.~\ref{Fig3}).
% bottom right).
These characteristics, namely reflection for $E_b$ large and
positive and the possibility of trapping, along with reflection
for $E_b$ large and negative, are illustrated in the
panels of Fig.~\ref{Fig4}.

\begin{figure}[tp]
\begin{center}
\includegraphics[width=0.6\textwidth]{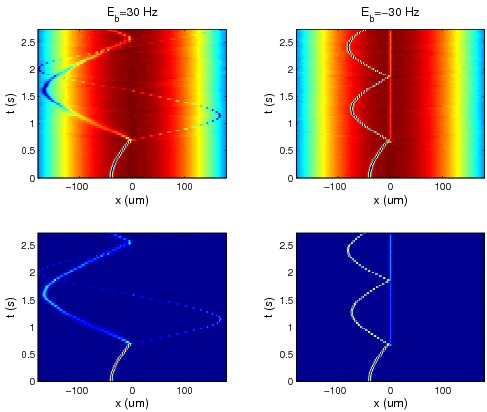}
\end{center}
\caption{(Color online) Similar to the previous two figures,
but for an intermediate value of $|E_b|$.}
%,  where it is not
%straightforward to reconstruct an effective potential.}
%between
%the case with repulsive bright impurity and the case with attractive
%bright impurity, being in both cases $E_d=0$. (Top row) the dark
%components \{$|u_n|^2$\} and (Bottom row) the bright components
%\{$|v_n|^2$\}.}
\label{Fig5}
\end{figure}

\begin{figure}[h]
\begin{center}
\includegraphics[width=0.7\textwidth]{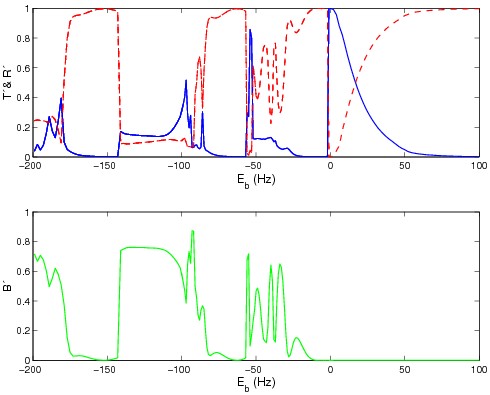}
\end{center}
\caption{\label{Fig10}(Top) Representation of $T'$ (full blue line, top panel) and $R'$ (dashed red line, top panel) and $B'$ (full green line, bottom panel) versus $E_b$ for the case $E_d=0$.}
\end{figure}

%It should also be highlighted that
%intermediate values of $E_b$ can be seen to interpolate
%
An example of intermediate values of $E_b$
and their associated dynamics can be found in Fig.~\ref{Fig5}.
In these examples, for positive $E_b$,
the impurity leads to partial transmission and
partial reflection, but does not enable the possibility
of trapping at the defect. The latter possibility
is explored for $E_b<0$, whereby there
is a fraction of atoms which is
trapped at the defect, while also a considerable
fraction appears to be reflected.
We notice that in the case
where there is no impurity affecting the dark component and
where there exists an attractive well in the bright one, as the well
depth $|E_b|$ increases, two regimes appear alternatively:
a reflection regime and a trapping-transmission-reflection regime (red
and brown colors, respectively, in region III of Fig.~\ref{Fig1}).
Indeed, this alternation may be quite complex and a characteristic
example thereof is presented in Fig.~\ref{Fig10},
%\jcm{
where the fractions of atoms transmitted through the defect, reflected from it and trapped
in the immediate vicinity of the barrier are measured. Those quantities are quantitatively
described by the transmission, reflection and trapping coefficients. Notice that the
time-dependent coefficients are defined as:
\begin{equation}
     T(t)=\frac{\int_\epsilon^\infty |v(x,t)|^2 dx}{\int_{-\infty}^\infty |v(x,t)|^2 dx}, \quad
     R(t)=\frac{\int_{-\infty}^{-\epsilon} |v(x,t)|^2 dx}{\int_{-\infty}^\infty |v(x,t)|^2 dx}, \quad
     B(t)=\frac{\int_{-\epsilon}^{\epsilon} |v(x,t)|^2 dx}{\int_{-\infty}^\infty |v(x,t)|^2 dx}.
\end{equation}
Figure~\ref{Fig10} depicts these coefficients
%We have represented in Fig.~\ref{Fig10} those coefficients
at time $t'$, namely $R'$, $T'$ and $B'$, with $t'$ being the time where the bright component of the DB
soliton reaches its maximum excursion. The figure illustrates that the full dynamics features more complicated
resonant type transmission events, as well as alternating windows
of predominantly reflection or predominantly trapping.
These complex scenarios are beyond the scope of the particle
analysis provided herein. We do note, however, the apparent similarity
of these results with the ones obtained in the case of a single
component bright soliton which scatters off of a quantum well~\cite{pantofl}.
In the latter case, the variational analysis was already fairly cumbersome
%for
even for a $\delta$-function potential, while here it is rendered
more elaborate by the presence of two-components and the Gaussian form
%nature
of the barrier. A more detailed analysis, perhaps in the
simpler $\delta$-function attractive setting based on a two-mode
variational ansatz would constitute an interesting problem for
further studies.
Nevertheless, we should point out that this behavior is similar to that obtained for the case where
there exists a repulsive barrier in the dark component and there is
no impurity affecting the bright component.

%It is interesting to observe that for small values of $|E_b|$ and
%$E_d=0$ a repulsive barrier in the bright component behaves as an
%attractive well and an attractive well in the bright component
%behaves as a repulsive barrier. Fig.~\ref{Fig3} illustrates the
%comparison of these two cases represented by IIA and IIIA,
%respectively, in Fig.~\ref{Fig1} with $E_b=10$ and $E_b=-10$. The
%top panels show the effective potentials $V_eff$. This
%counter-intuitive result is in accordance with ~\cite{Achi}.

%Let us compare also high values of $|E_b|$ when $E_d=0$ (IIB and
%IIIB of Fig.~\ref{Fig1}). Fig.~\ref{Fig4} shows the effective
%potentials and the contour plots of the densities of the dark and
%bright components for $E_b=50$ (column IIB) and $E_b=-50$ (column
%IIIB).

%Fig.~\ref{Fig5} illustrates the contour plots of the densities of
%the dark and bright components for an intermediate value of $|E_b|$
%with $E_d=0$. In the left column we consider the value $E_b=30$ and
%in the right column the value $E_b=-30$.

%\section{SEPARATE FIGURES JUST IN CASE.....?????????????????}

%We have studied the evolution of the transmission (T) and reflection
%(R) rates of the dark-bright components, being defined as:

%\begin{equation}
%     T(t)=\frac{\int_0^\infty |v(x,t)|^2 dx}{\int_{-\infty}^\infty
% |v(x,t)|^2 dx},
%\end{equation}

%\begin{equation}
%     R(t)=\frac{\int_{-\infty}^0 |v(x,t)|^2 dx}{\int_{-\infty}^\infty
% |v(x,t)|^2 dx},
%\end{equation}

\section{Conclusions and Future Challenges}

In the present work, we have presented a survey of the collisions of
dark-bright (DB) solitons with defects. Initially motivated by
the potential of experimental studies (a prototypical example
of which was shown herein), we considered the setting of
a dark-bright soliton impinging on a defect potential.
In the case of a bright soliton hitting a well~\cite{pantofl,pantofl1}
or a barrier~\cite{gardiner,martin}, this theme has been of
intense theoretical and even experimental~\cite{randy,dic}
interest recently. However, far less has been done in
the realm of dark-bright solitons.

We have shown
%revealed
that in the case of two equal repulsive barriers acting on both components,
%leads
the DB solitons demonstrate
%to
a clear classical particle behavior, which
%phenomenology, involving
involves transmission
for weak potentials and reflection for strong ones. Similarly,
predominantly transmission type events were observed
for equal attractive potentials acting in both components.

On the other hand, we illustrated that more complex scenarios
can develop in the case where the impurity acts only
on one of the two components. We categorized these cases,
illustrating the analogies of a repulsive barrier in the first
component with an attractive one in the second component (and vice-versa).
We explained the low barrier amplitude cases on the basis of
somewhat counter-intuitive, cross-component effective potentials
and argued that the large amplitude cases may be significantly
different due to the role of the defect in both components.
We showcased the complexity of the latter by means of
cases containing transmission and reflection, or trapping,
transmission and reflection together and by monitoring
the dependence of the different fractions (of trapping, transmission
or reflection), as a function of the barrier amplitude.

It would certainly be interesting to extend this chiefly numerical
(but also experimental) study further. On the experimental side,
it would be extremely interesting, although more challenging,
to engineer potentials that are selective to particular hyperfine
states, so that some of the predictions proposed herein could
be tested. From a theoretical perspective, it would be very
relevant to attempt to distill a simple setting (e.g.
a $\delta$-function potential) where a theoretical study of
the above reported phenomenology could be appreciated in
more quantitative terms. 
Numerically, it may also be quite significant to appreciate
the effect of the width of the barrier, as here we have concentrated
on the sign and magnitude (and inter-component interplay of the) barrier.
Natural extensions may also concern
the possibility of scattering in higher-dimensional
settings and evaluation of the role of transverse degrees of
freedom therein.

{\bf Acknowledgments.} P.E. acknowledges financial support from NSF and ARO.
P.G.K. gratefully acknowledges support from
NSF-DMS-0806762 and the Alexander von
Humboldt Foundation, as well as the Binational Science Foundation.
A.A., F.R.R. and J.C. acknowledge financial support
from the MICINN project FIS2008-04848. The work of D.J.F. was partially supported by the
Special Account for Research Grants of the University of Athens.

\end{document}